\documentclass[11pt,superscriptaddress,aps,prd,preprint]{revtex4}
\usepackage[active]{srcltx}
\usepackage{graphicx}
\usepackage[utf8]{inputenc}
\usepackage{amsmath}
\usepackage{xcolor}
\usepackage{times,txfonts}
\usepackage{float}%para fixar as figuras na posição desejada

\begin{document}

\title{Black Strings in Asymptotically Safe Gravity}

\author{M. Nilton}
\affiliation{Universidade Federal do Cear\'{a} Campus de Russas, Russas, Cear\'{a}, Brazil}
\email{matheus.nilton@fisica.ufc.br}

\author{G. Alencar}
\affiliation{Universidade Federal do Cear\'{a}, Fortaleza, Cear\'{a}, Brazil}
\email{geova@fisica.ufc.br}

\author{R. N. Costa Filho}
\affiliation{Universidade Federal do Cear\'{a}, Fortaleza, Cear\'{a}, Brazil}
\email{rai@fisica.ufc.br}

\date{\today}

\begin{abstract}
In this paper, we study black strings in asymptotic safety gravity (ASG) scenario. The ASG approach is introduced by implementing gravitational and cosmological running coupling constants directly in the black string metric. We calculate the Hawking temperature, entropy, and heat capacity of the improved black string metric in two cases: considering the cosmological constant fixed in some fixed point and the general case where both Newton's constant and cosmological constant are improved. For the identification of the scale moment we used an general inverse law setting $k(r)\sim 1/r^{n}$. We show that improving only the Newton's constant the problem of singularity is solved for the identifications with $n>1$. However, if the cosmological constant is also running the singularity persists in the solution. Also, we show that the ASG effects predicts the presence of a remnant mass in the final evaporation process. Besides that, a logarithmic correction is observed in the entropy. However, a running cosmological constant introduces new correction terms to the entropy beyond that. We show that the improved black string solution remains stable, as in the usual case. Phase transitions are not observed in both cases studied here.

\end{abstract}

\maketitle

\section{Introduction}

The research in black hole physics has received renewed interest in the last few years. This is mainly due to the present era of high precision measurements, which allowed, for example,  the direct measurement of gravitational waves by LIGO  and the image of supermassive black holes by EHT \cite{LIGOScientific:2016aoc,EventHorizonTelescope:2019dse}. However, the black hole interior is riddled with a spacetime singularity. To better understand the properties of these singularities, it is interesting to study their cylindrical counterparts, called black strings. 

Black strings/black branes are solutions of the Einstein field equations in D-dimensions \cite{Duff:1987cs}. These solutions can be considered generalizations of black holes with translational symmetry. However, contrary to the spherically symmetric counterpart, black strings just exist in asymptotically anti-de-Sitter (AdS)\cite{Bellucci:2010gb}. In particular, the AdS spacetime is important due to the AdS/CFT conjecture \cite{Becker:2006dvp}. They can be used to test the duality of this theory and gives an alternative to understanding the physics of gauge field theories \cite{Witten:1998qj,Witten:1998zw,Henningson:1998gx,Polchinski:1995mt,Polchinski:1994fq}.
 In particular, a solution of black string in the usual four-dimensional spacetime in cylindrical coordinates was found by Lemos in \cite{Lemos:1994xp}. Since then, these black string solutions  have been used in several works \cite{Carvalho:2022eli,Henriquez-Baez:2022ubu,Muniz:2022vhk,Muniz:2022otq,Sriling:2021lpr,Awad:2002cz}. 

A subject of great importance is the interaction of the quantum nature of particles and geometry. When we take these interactions into account, new phenomena can be predicted, such as the radiation emitted by black holes \cite{Hawking:1975vcx}. However, when we describe quantum gravity along the lines of quantum field theory, we obtain a non-renormalizable theory. One way of implementing quantum corrections in general relativity is the asymptotic safety conjecture proposed by Weinberg \cite{Weinberg}. This conjecture states that a theory with non-Gaussian fixed points, such that all the running coupling constants of the theory tends to them in the ultraviolet (UV) limit, is free of divergences and, therefore, is renormalizable and predictive. The search for fixed points in the case of gravitational interaction has been done \cite{Reuter:1996cp, Lauscher:2001ya, Litim:2003vp, Machado:2007ea, Manrique:2011jc, Christiansen:2012rx, Morris:2015oca, Demmel:2015oqa, Platania:2017djo, Christiansen:2017bsy, Falls:2018ylp, Narain:2009fy, Oda:2015sma,Draper:2020bop,Platania:2020knd,Fehre:2021eob}. 

In the asymptotic safety approach, we have to compute the gravitational flow $\Gamma_k[g_{\mu\nu}$], which is the functional of the metric and satisfies the exact renormalization group equation (ERGE) \cite{Reuter:1996cp}
\begin{equation}
k\partial_k\Gamma_k=\frac{1}{2}\textrm{Tr}[(\Gamma^{(2)}_k+\mathcal{R}_k)^{-1}k\partial_k\mathcal{R}_k],
\end{equation}
where $k$ is the parameter of the renormalization group, $\Gamma^{(2)}_k$ is the Hessian of $\Gamma_k[g_{\mu\nu}]$ and $\mathcal{R}_k$ is the cutoff function, responsible for eliminating the small moment modes, that is, the infrared (IR) divergences. Due to the difficulty to solve the ERGE, we have to use the truncation methods to obtain approximate solutions for $\Gamma_k$. The most natural truncation consists in an expansion of $\Gamma_k$ in the basis $\sqrt{g}$ and $\sqrt{g}R$, which is equivalent to associating the gravitational flow with the Einstein-Hilbert action. Therefore this method is called Einstein-Hilbert truncation.  However, it is possible to consider other terms beyond the Ricci scalar in the asymptotic safety approach. With this, the fundamental constants of the theory become a function of the renormalization group parameter $k$, and their form is determined in each method cited above. 

In general, the quantum corrections from these methods can be studied by putting the running coupling constants directly in the classical solutions. Then, the properties of the improved solutions can be done. 
This strategy was used to study several gravitational systems in four dimensions, such as black holes \cite{Saueressig:2015xua,Pawlowski:2018swz,Platania:2019kyx,Bosma:2019aiu,Ishibashi:2021kmf,Chen:2022xjk} and wormholes \cite{Moti:2020whf,Nilton:2022cho,Alencar:2021enh, Alencar:2021mus,Nilton:2022hrp}. The consequences for the BTZ  three-dimensional counterpart of black holes have also been extensively studied  in Refs. \cite{Koch:2016uso,Rincon:2017ypd,Rincon:2017goj,Rincon:2018sgd,Rincon:2018dsq,Rincon:2018lyd,Rincon:2019zxk,Fathi:2019jid,Rincon:2022hpy}. However, up to now, no study about black strings has been done about improved black strings. 

This is the first in a series of papers where we study black strings in the asymptotic safety gravity context. It is organized as follows: in the next section, we improve on the black string metric. In section III we compute and analyze the Hawking temperature, entropy, and heat capacity of the improved black string solution. In section IV we conclude the paper.

\hspace{0.4cm}

\section{Improved Black String Metric in Asymptotic Safety Gravity}

Let us consider the static case of a neutral black string solution in the usual four-dimensional spacetime reported in \cite{Lemos:1994xp}. The metric has the following form
\begin{equation}
\label{bs-metric}
ds^2=-f(r)dt^2+\frac{dr^2}{f(r)}+r^2\,d\varphi^2+\alpha^2 r^2 dz^2,
\end{equation}
where we use cylindrical coordinates $(t,r,\varphi,z)$ with the ranges $-\infty<t<\infty$, $0\leq r<\infty$, $0\leq\varphi<2\pi$  and $-\infty<z<\infty$. Here, $\alpha$ is a parameter that is related to 
the cosmological constant $\Lambda$ through $\alpha^2 = -\Lambda/3>0$, since the 4D black string is a solution of Einstein's field equations with a cosmological constant. The function $f(r)$ is given by
\begin{equation}\label{blackstring}
f(r)=\alpha^2 r^2 - \frac{b}{\alpha r},
\end{equation}
where $b$ is a constant that is proportional to the mass of the black string. Although the suitable name for the object described by the metric \eqref{bs-metric} is black string, this object can be interpreted as a cylindrical black hole or a black brane. 

The black string is not a regular solution. Indeed, this is manifested in the curvature scalars, such as the Kretschmann scalar, which for the classical black string metric is given by $K=R_{\mu\nu\sigma\lambda}R^{\mu\nu\sigma\lambda}=24\alpha^4\left(1+\frac{b^2}{2\alpha^6 r^6}\right)$, which diverges at the origin. Another interesting fact is that the Ricci scalar will be constant and proportional to the cosmological constant given by $R=4\Lambda_0=-12\alpha^2$, that is, in the spatial infinity the curvature will not be zero. 

%This is a solution to Einstein's field equations of general relativity with a cosmological constant in cylindrical spacetime. The cylindrical and temporal symmetries allow the following form for the line element of a black string \cite{Lemos:1994xp}.
%\begin{equation}
%\label{bs-metric}
%ds^2 = -f(r)dt^2+\frac{dr^2}{f(r)}+r^2 d\varphi^2+\alpha^2 r^2 dz^2,
%\end{equation}
%where we use cylindrical coordinates $(t,r,\varphi,z)$ with the ranges $-\infty < t < \infty$, $0\leq r < \infty$, $0\leq\varphi< 2\pi$ and $-\infty<z<\infty$. The constant $\alpha$ is defined by $\alpha^2\equiv -\Lambda/3 > 0$, where $\Lambda$ is the cosmological constant. The function $f(r)$ for the classical black string solution is given by
%\begin{equation}
%f(r)=\alpha^2 r^2 - \frac{4\mu}{\alpha r},
%\end{equation}
%where $\mu$ is the mass per unit length. Although the suitable name for the object described by the metric \eqref{bs-metric} is black string, in a 4D spacetime this object can be interpreted as a cylindrical black hole. 

To consider quantum corrections in the black string metric \eqref{bs-metric}, we will improve the fundamental coupling constants of the theory, turning it into running coupling constants, that is
\begin{equation}
\Lambda \rightarrow \Lambda(k), \,\,\,\, G \rightarrow G(k),
\end{equation}
where $k$ is the scale moment of the renormalization group method and determines the value of the running constants on each energy scale. The scale moment $k$ must be identified with a function of the distance that dictates the behavior of the coupling constants for the various energy scales, and there are several ways to do that. Among the possibilities, the scale $k$ can be identified with physical quantities of interest in gravitational systems, such as the geodesic distance or curvature invariants constructed with the components of the Riemann tensor. In this paper, for simplicity, we consider an inverse distance law for setting the scale moment, i.e, $k(r)=\zeta/r^n$, where $\zeta$ is an unknown constant and $n$ is a positive integer. This simple identification ensures that in the IR limit, the quantum corrections become negligible. A particular kind of this identification was used to study the physical properties of the quantum-improved BTZ black holes in Refs \cite{Koch:2016uso,Rincon:2017ypd,Rincon:2017goj,Rincon:2018sgd,Rincon:2018dsq,Rincon:2018lyd,Rincon:2019zxk,Fathi:2019jid,Rincon:2022hpy}.

For the running coupling constants, we will use the analytical expressions for the improvements obtained by Koch and  Ramirez in a seminal paper, and is given by\cite{Koch:2010nn}
\begin{eqnarray}
g(k)&=&\frac{k^2}{1+k^2/g^*} \quad,\label{gk}\\
 \lambda(g)&=&\frac{g^*\lambda^*}{g}
\left((5+e)\left[1-g/g^*\right]^{3/2}-5+3g/(2g^*)(5-
g/g^*)\right) \quad,\label{lvong}
\end{eqnarray}
where
$
e=\Lambda_0/(g_*\lambda_*),
$
with $g_*=0.707$ and $\lambda_*=0.193$ being the non-Gaussian fixed points and $\Lambda_0$ the IR value of the cosmological constant. These expressions are written in terms of the dimensionless running constants $g(k)=k^{2}G(k)$ and $\lambda(k)=k^{-2} \Lambda(k)$ in units where $G_0=1$. With the above expression, we can analyze all the properties of our solution, including thermodynamics. 

Following other authors \cite{Bonanno:2000ep,Adeifeoba:2018ydh}, we will implement the quantum corrections due to asymptotic safety formalism putting the expressions \eqref{gk} and \eqref{lvong} directly in the metric. Therefore, the quantum-improved black string remains static with cylindrical symmetry, but the lapse function \eqref{blackstring} is changed to
\begin{equation}
\label{improved-f}
f(r)=\alpha^2(k) r^2 - \frac{G(k) b}{\alpha(k) r},
\end{equation}
where $\alpha^2(k)=-\Lambda(k)/3$. 

In what follows, we study the thermodynamic properties of the improved metric in two cases: considering that the cosmological constant is already at a fixed point, where only Newton's constant becomes running, and the case where both coupling constants are dependent on the scale moment.

\section{Curvature and thermodynamics of the improved black string with the cosmological constant fixed}

Now, we will check how ASG corrections modify the thermodynamic quantities of the black string. Therefore, in order to analyze the physics of the black string dependent on the scale of energy we will compute the temperature, entropy, and heat capacity using the improved metric with the lapse function \eqref{improved-f}. Following other authors \cite{Chen:2022xjk}, for simplicity, we will firstly consider the case where the cosmological constant is already at a fixed point, and then only Newton's constant becomes a function of the distance, given by:
\begin{equation}
G(r) = \left(1+\frac{\xi}{r^{2n}}\right)^{-1},
\end{equation}
where we use the identification $k(r)=\zeta/r^n$ and defined $\xi=\zeta^2 g_*^{-1}$. The lapse function turns into
\begin{equation}
\label{improved-f-1}
f(r)=\alpha^2 r^2 - \frac{b}{\alpha r}\left(1+\frac{\xi}{r^{2n}}\right)^{-1}.
\end{equation}

It is interesting to determine how the introduction of a running Newton's constant modify the curvature of the black string, given by the metric \eqref{bs-metric} with the lapse function \eqref{improved-f-1}. In this case, the Ricci and Kretschmann scalars will be given respectively by the expressions
\begin{equation}
R = -\frac{2(6r^{3+6n}\alpha^3 +nb(2n-1)r^{4n}\xi+18r^{3+4n}\alpha^3 \xi -nb(2n+1)r^{2n}\xi^2 +18r^{3+2n}\alpha^3 \xi^2 + 6r^3 \alpha^3 \xi^3)}{r^3 \alpha (r^{2n}+\xi)^3},
\end{equation}
and
\begin{eqnarray}
K &=& 24\alpha^4 - \frac{8nbr^{-3+2n}\alpha\xi ((1-2n)r^{2n
}+\xi+2n\xi)}{(r^{2n}+\xi)^{3}} + (4b^2 r^{-6+4n}(3r^{8n}-2(2n^2+5n-2)r^{6n}\xi \nonumber\\
&&+(4n^4 +12n^3 +9n^2-30n+18)r^{4n}\xi^2-2(4n^4 - 15n^2 + 15n -6)r^{2n}\xi^3+(4n^4 - 12n^3 +17n^2 - 10n + 3)\xi^4))\times\nonumber\\
&&\times(\alpha^{-2} (r^{2n}+\xi)^{-6}).
\end{eqnarray}

\begin{figure}[!htpb]
    \centering
    \includegraphics{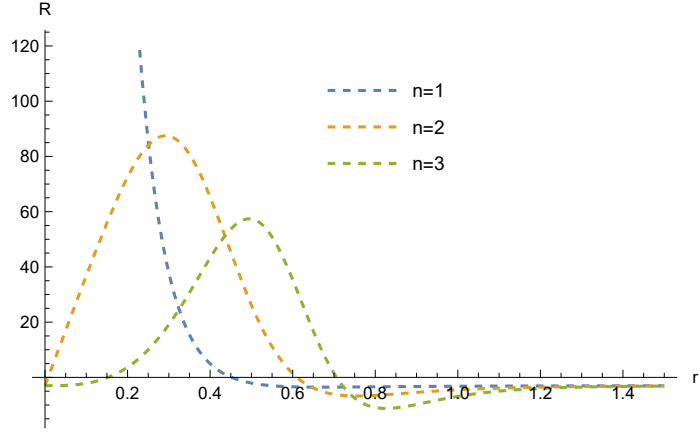}
    \caption{Ricci scalar for the black string metric for Newton's constant running case. We considered $\xi=0.10$, $\alpha=0.5$ and $b=1$. Note that the divergence occurs only for the case where $n=1$. For $n>1$ the curvature is well }
    \label{g-ricciscalar}
\end{figure}

\begin{figure}[!htpb]
    \centering
    \includegraphics{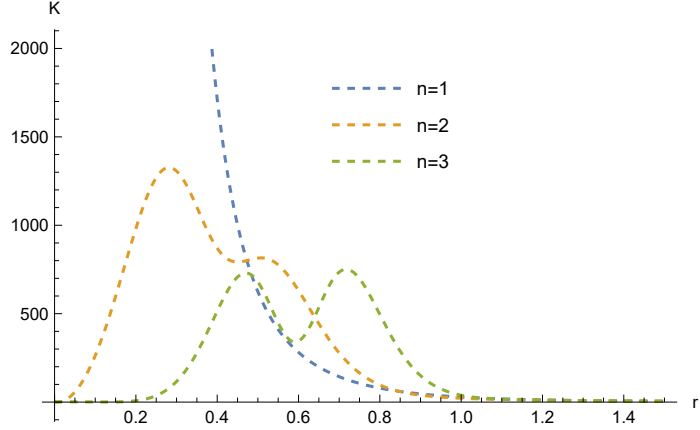}
    \caption{Kretschmann scalar for the black string metric for Newton's constant running case. We considered $\xi=0.10$, $\alpha=0.5$ and $b=1$.}
    \label{g-kretschmannscalar}
\end{figure}

We can see that both expressions for $R$ and $K$ tends to their classical expressions for $\xi\rightarrow 0$ and in the limit of great distances. Only for the case $n=1$ the introduction of a running Newton's constant aggravates the singularity problem. Indeed, for this case we can see that not only $K$ presents a singularity, but $R$ also presents the singularity in the origin. However, the order of singularity is smoothed: for $r\approx0$ the Ricci scalar will diverge with $1/r$ and the Kretschmann scalar with $1/r^2$, while in the classical theory, the Kretschmann scalar diverges with $1/r^6$ but the Ricci scalar will be constant. In contrast, for the case $n>1$ both $R$ and $K$ are regular in the origin. Therefore, the improved black string solution in this case is geodesically complete, and the problem of the singularity is cured. Although that, for $n>1$ there will be a value of the radial coordinate where the curvature becomes a maximum. In Figs.(\ref{g-ricciscalar}) and (\ref{g-kretschmannscalar}) we can see that the behavior of the Ricci scalar and the Kretschmann scalar will be essentially the same.

The Hawking temperature is defined by 
\begin{equation}
\label{temperature}
T_{\textrm{H}}(r_+)=\frac{\left. \partial_r f(r)\right|_{r=r_+}}{4\pi},
\end{equation}
where $r_+$ is the horizon radius of the black string (remember that the black string can be interpreted as a cylindrical black hole). The horizon radius is defined as $f(r_+)=0$, and from the Eq.\eqref{improved-f-1} we obtain the mass of the black string in function of the horizon radius
\begin{equation}
\label{mass-function}
b(r_+)=\alpha^3 r_+^{3-2n}(r_+^{2n}+\xi)
\end{equation}

\begin{figure}[!htpb]
    \centering
    \includegraphics{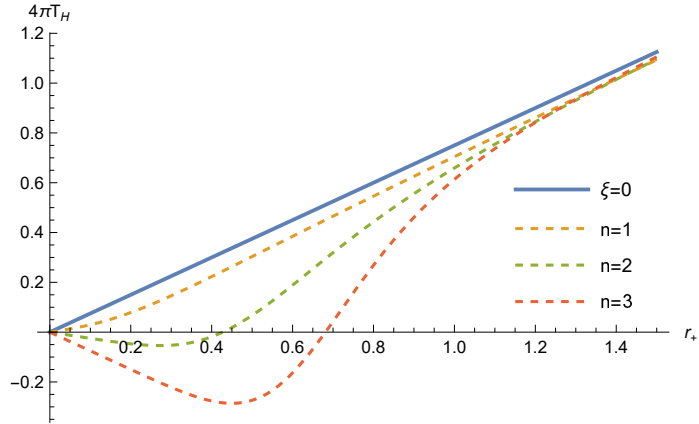}
    \caption{Hawking temperature improving only the Newton's constant in the black string metric. The thick line is the usual temperature, where $\xi=0$. The dashed lines are for the temperature with quantum gravity corrections. We have used $\xi=0.10$ and $\alpha=0.5$.}
    \label{temperature-bs-asg}
\end{figure}

Calculating the expression \eqref{temperature} we get for the temperature of the improved black string as a function of the horizon radius
\begin{equation}
\label{g-temperature}
4\pi T_{\textrm{H}}(r_+)=\alpha^2 r_+\left(3-\frac{2n\xi}{r_+^{2n}+\xi}\right).
\end{equation}
The above expression tends to the usual expression for the temperature of the black string both in the limit $r_+>>0$ and when $\xi\rightarrow0$, where the ASG corrections become irrelevant. The behavior of the temperature for the cases $n=1$, $n=2$ and $n=3$ can be seen in Fig.(\ref{temperature-bs-asg}). We can note that the deviation from the linearity of the usual expression for the temperature of the black string becomes relevant only for small values of $r_+$, as we can expect for quantum gravity corrections. Besides that, as we increase the value of $\xi$, there is a slight decrease in temperature. However, only for the case $n=1$ the temperature of the improved black string metric converges to zero when the horizon radius becomes null, such as the usual case, and therefore the identification $k(r) \propto 1/r$ does not predict the presence of a remnant mass. In contrast, as we can see in the plot of the temperature, the cases $n>1$ ensures the presence of a remnant mass in the final of the evaporation process of the black string. For $n>1$ the radius of the remnant is given by
\begin{equation}
r_{\textrm{rem}}=\sqrt[2n]{\frac{2n\xi}{3}-\xi}.    
\end{equation}

\begin{figure}[!htpb]
    \centering
    \includegraphics{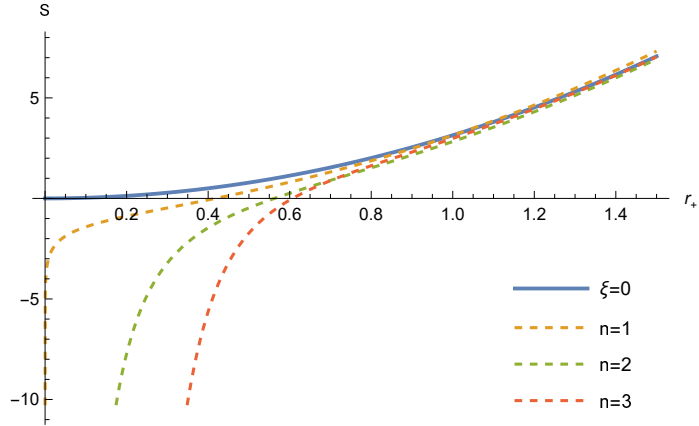}
    \caption{Effects of a running Newton's constant in entropy of  the black string metric. The thick line is the usual temperature, where $\xi=0$. The dashed lines are for the entropy with quantum gravity corrections. We have used $\xi=0.10$ and $\alpha=0.5$.}
    \label{entropy-bs-asg}
\end{figure}

Now, we compute the entropy for the improved black string metric \eqref{improved-f-1}. For this, we can use the first law of the black hole thermodynamics $dM=T_{\textrm{H}}\,dS$. The only difference is that the parameter associated to the mass is $b$, being proportional to the linear density of mass of the black string. Therefore, we compute $d\,b=T_{\textrm{H
}}d\mathcal{S}$ where $\mathcal{S}$ is proportional to the linear density of entropy. Using the Eqs. \eqref{mass-function},\eqref{g-temperature} we obtain for entropy
\begin{equation}
\mathcal{S}(r_+)= 
\left\{
\begin{array}{cc}
2\pi\alpha(r_+^2+2\xi\ln{r_+}), \quad\quad\textrm{for}\quad\quad n=1,\\
\frac{2\pi r_+^2\alpha(n-1-r_+^{-2n}\xi)}{n-1}\quad\quad\textrm{for}\quad\quad n>1.
\end{array}
\right.
\end{equation}
Clearly, in the limit $\xi\rightarrow 0$, when the ASG corrections are irrelevant, we obtain the usual behavior for the entropy of a black string, i.e., a quadratic function of the horizon radius $r_+$. We can note that the ASG brings a logarithmic correction to the entropy only for the case $n=1$. This feature is similar to what happens to other quantum gravity approaches, such as the generalized uncertainty principle (GUP) and noncommutative geometry \cite{Medved:2005vw,Chen:2009sp,Banerjee:2010qc,Sen:2012dw,Bagchi:2013qva,Keeler:2014bra,Nozari:2006vn,Faizal:2014tea,Anacleto:2020zfh}. Although that, the logarithmic correction of the entropy is also observed in other asymptotic anti-de Sitter spacetimes, such as Schwarzschild-anti-de Sitter black holes \cite{Ma:2014zia}. But, for the identifications with $n>1$, the correction for the entropy comes from the inverse of the horizon radius, emphasizing that the ASG corrections becomes relevant only for small distances. The behavior of the entropy for the cases $n=1$, $n=2$ and $n=3$ in comparison with the usual case can be seen in Fig.(\ref{entropy-bs-asg}).

\begin{figure}[!htpb]
    \centering
    \includegraphics{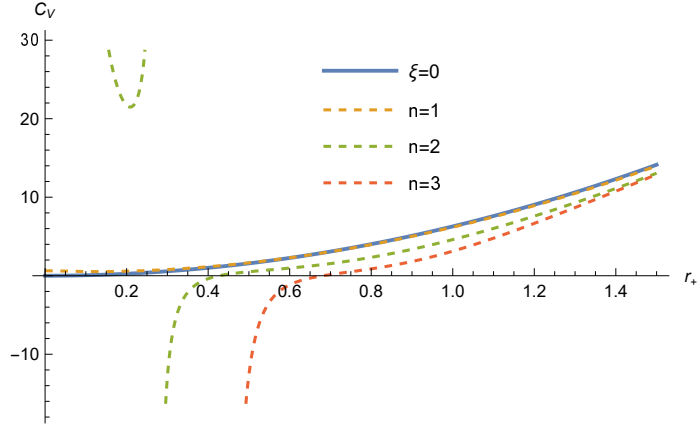}
    \caption{Heat capacity for the case where only Newton's constant is running in the black string metric. The thick line is for $\xi=0$, i.e., without ASG corrections. The dashed lines is for the $\xi=0.10$ (orange lines) and for $\xi=0.15$ (green line). In all lines, we have set $\alpha=0.5$.}
    \label{heat-capacity}
\end{figure}

Finally, let us determine the heat capacity at constant volume $C_V$, defined by 
\begin{equation}
\label{def:heat-capacity}
C_V(r_+)=\frac{\partial M}{\partial T_\textrm{H}}=\frac{\partial M}{\partial r_+}\left(\frac{\partial T_{\textrm{H
}}}{\partial r_+}\right)^{-1}.
\end{equation}
Using the Eqs.\eqref{mass-function} and \eqref{g-temperature} we obtain the following expression for the heat capacity (more precisely, if we use $b(r_+)$ we find a linear density of heat capacity, but the behavior is the same for the heat capacity itself):
\begin{equation}
C_V=\frac{4\pi\alpha r^{2-2n}(r^{2n}+\xi)^2 (3r^{2n}+(3-2n)\xi)}{3r^{4n}+2(3+n(2n-1))r^{2n}\xi+(3-2n)\xi^2}.
\end{equation}
Once again, we can see that the heat capacity tends to its usual behavior both for large values of $r_+$, and in the limit $\xi\rightarrow 0$, where the ASG corrections vanish. In fig.(\ref{heat-capacity}) we plotted the behavior of the heat capacity for $n=1$, $n=2$ and $n=3$ cases in comparison to the usual expression. The heat capacity for the improved black string metric is always positive and therefore remains a stable system, such as the usual case obtained by the classical metric. The negative region for the cases $n>1$ is for $r_+<r_{\textrm{rem}}$ and therefore, the improved heat capacity becomes negative only when the temperature is negative too, being a forbidden region. Contrary to what is observed with the other thermodynamic quantities, the heat capacity increases when $\xi$ increases. Besides that, we can see that the evaporation process ends when $r_+ = r_{\textrm{rem}}$, and therefore, the ASG corrections predicts the existence of a remnant mass only for $n>1$ cases, as already noted from the improved temperature expression. Once more time, the divergences of the heat capacity in the cases $n>1$ happens only when $r_+<r_{\textrm{rem}}$, indicating that we do not have phase transitions both in the usual and in the corrected black string solution. 

\section{Curvature and thermodynamics of the improved black string with a running the cosmological constant}

Now, we will determine how the inclusion of both running Newton's constant and the cosmological constant affects the thermodynamic properties of the black string. Following the previous procedure, using Eqs. \eqref{gk} and \eqref{lvong} the improved lapse function $f(r)$ with the scale identification $k=\zeta/r^{n}$ provides:
\begin{eqnarray}
f(r)&=-\frac{g_{*} \lambda_{*} r^2}{3}\left(1+\frac{\zeta^2}{g_{*} r^{2n}}\right)\left(-5+(5+e)\left(\frac{g_{*} }{g_{*} +\zeta^2 r^{2n}}\right)^{3/2}+\frac{3}{8+10g_{*} r^{2n}\zeta^{-2}}\right)\nonumber\\
&-\frac{\sqrt{3}g_{*} b }{r(g_{*} +\zeta^2 r^{-2n})\sqrt{-g_{*} \lambda_{*} \left(1+\frac{\zeta^2}{g_{*} r^{2n}}\right)\left(-5+(5+e)\left(\frac{g_{*}}{g_{*} +\zeta^2 r^{-2n}}\right)^{3/2}+\frac{3}{8+10g_{*}\zeta^{-2}r^{2n}}\right)}} \label{lf}.
\end{eqnarray}
The above function tends to the usual black string metric in the limit $\zeta\rightarrow 0$, as can be easily checked. 

As we did for the previous case, it is instructive to analyze the behavior of the curvature invariants in this case too. Due to the long length of the expressions for the Ricci scalar $R$ and the Kretschmann scalar $K$ for this case, we show the behavior of these scalars only graphically. 

\begin{figure}[!htpb]
    \centering
    \includegraphics{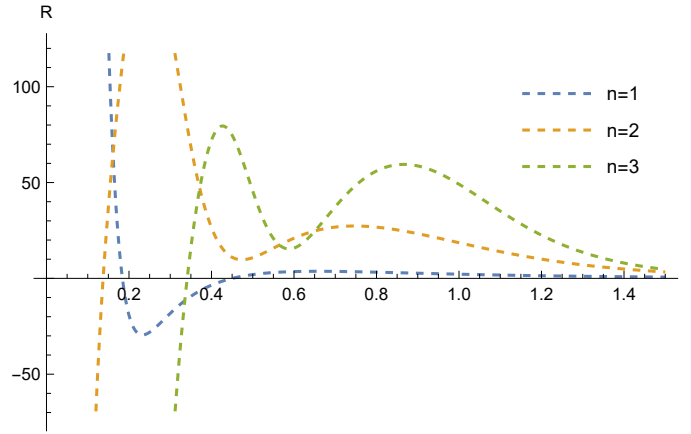}
    \caption{Ricci scalar for the black string metric for the general case, where both constants are improved. We considered $\xi=0.10$, $b=1$ and $e=-0.10$.}
    \label{l-ricciscalar}
\end{figure}

The plot of the Ricci scalar can be seen in Fig.(\ref{l-ricciscalar}). We can see that the behavior of the Ricci scalar is very different from its classical counterpart, which is constant for all values of $r$. The improvements of the coupling constants imply in regions where the curvature is negative and positive. We can note that there will be a point where the Ricci scalar will be a maximum. However, contrary to what happens in the case where only Newton's constant is improved, the Ricci scalar will diverge at the origin. This leads to the interesting result that the singularity problem is solved only when just the Newton's constant is improved.

\begin{figure}[!htpb]
    \centering
    \includegraphics{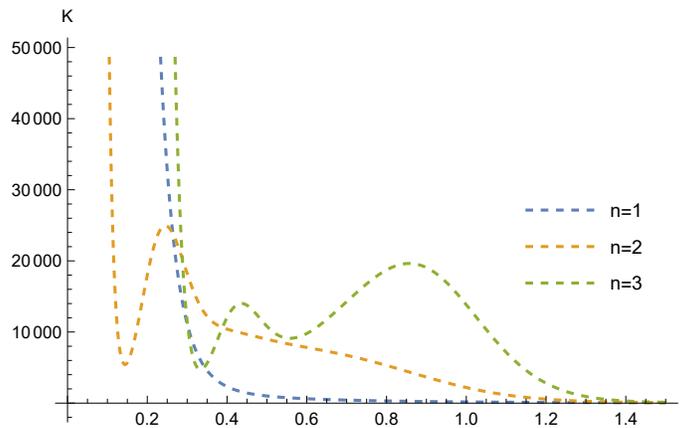}
    \caption{Kretschmann scalar for the black string metric for the general case case. We considered $\xi=0.10$, $b=1$ and $e=-0.10$.}
    \label{l-kretschmann}
\end{figure}

In Fig.(\ref{l-kretschmann}) we can see the behavior of the Kretschmann scalar for the improved black string metric in this general case. We can see that the ASG corrections provide a Kretschmann scalar with similar behavior to Ricci scalar. However, just as happens for the Ricci scalar, there will be a divergence in the origin, and therefore, the singularity problem is not solved in this case.  

Returning to thermodynamic, the horizon condition $f(r_+)=0$ allow us to write the mass parameter $b$ in function of the horizon radius
\begin{equation}
\label{lb}
b(r_+)=\frac{r_+^3 (g_* + r_+^{-2n}\zeta^2)}{3\sqrt{3}g_*}\left(-g_{*}\lambda_{*}\left(1+\frac{\zeta^{2}}{g_{*}r_+^{2n}}\right)\left(-5+(5+e)\left(\frac{g_{*}}{g_{*}+\zeta^2 r_+^{-2n}}\right)^{3/2}+\frac{3}{8+10g_* r_+^{2n}\zeta^{-2}}\right)\right)^{3/2}.
\end{equation}
In this way, making use of the Eqs.\eqref{lf} and \eqref{lb} we can compute the Hawking temperature through the definition \eqref{temperature}. With a direct calculation we can show that the Hawking temperature for the improved black string metric in the case where the cosmological constant is also running is given by
\begin{eqnarray}
4\pi T_{\textrm{H}}(r_+)&=&\frac{\lambda_{*} r_+}{6}((\left(\frac{g_{*}}{g_{*}+\zeta^2 r^{-2n}}\right)^{3/2}(-30g_{*}-6eg_{*}-30\zeta^2 r_+^{-2n}-6e\zeta^2 r^{-2n}+5n\zeta^2 r^{-2n}+en\zeta^2 r^{-2n})\nonumber\\
&&+(5g_{*}r_+^{2n}+4\zeta^2)^{-2}(750g_{*}^{3}r_+^{4n}+1905g_{*}^{2}\zeta^2 r_+^{2n}-1205ng_{*}^2\zeta^2 r_+^{2n}+1599g_* \zeta^4 -1880n g_* \zeta^4 \nonumber\\
&&+444r^{-2n}\zeta^6 - 740n r^{-2n}\zeta^6)) \label{l-temperature}.
\end{eqnarray}
One more time we can check that this expression tends to the behavior of the temperature obtained using the classical black string metric in the IR limit and for $\zeta\rightarrow0$, that is, in these limits we obtain a linear function of the radial distance. 

\begin{figure}[!htpb]
    \centering
    \includegraphics{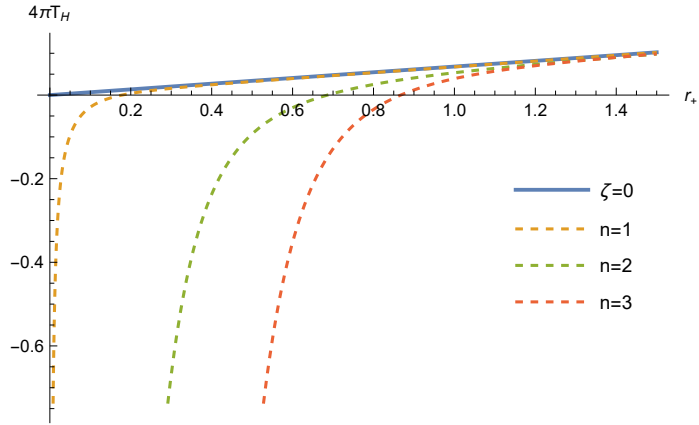}
    \caption{Hawking temperature for the improved black string in comparison to the temperature obtained from the classical metric ($\zeta=0$). We have set $\zeta=0.10$ and $e=-0.10$ in the plots of the improved solutions.}
    \label{l-temperature-plot}
\end{figure}

In Fig.(\ref{l-temperature-plot}) we show the behavior of the temperature for the improved black string spacetime. As expected, we can see that the deviation from the linearity occurs for small values of the horizon radius $r_+$. Besides that, if we increase the value of the parameter $\zeta$, meaning that the ASG effects becomes more relevant, the temperature becomes smaller. Contrary to what happens in the case where only Newton's constant is running, we can observe that the temperature has a zero for a non-null value of the horizon radius $r_+$ also for the case $n=1$. This means that the running cosmological constant introduces the presence of a remnant mass for all identifications of the type $k \propto 1/r^{n}$. The increase of $n$ just leads to an increase of the radius of the remnant mass. Indeed, we always have the presence of a remnant in this case. From the temperature expression, we can obtain the equation that determines the radius of the remnant mass. Defining $u=r_+^{2n}$, the resulting equation will be a polynomial of degree $9$ with the form
\begin{eqnarray}
a_0 + a_1 u + a_2 u^2 + a_3 u^3 + a_4 u^4 + a_5 u^5 + a_6 u^6 + a_7 u^7 + a_8 u^8 + a_9 u^9 = h(u),
\end{eqnarray}
whose coefficients are given by
\begin{eqnarray}
a_0 &=& -21904 (3 - 5 n)^2 \zeta^{18}, \quad\quad a_1 = -1480 (-3 + 5 n) (-453 + 598 n) g_* \zeta^{16}, \nonumber\\
a_2 &=&-15 (606639 + 8 n (-199841 + 127565 n)) g_*^2 \zeta^{14}, \nonumber\\
a_3 &=& (2560 e (-6 + n)^2 + 256 e^2 (-6 + n)^2 - 
   5 (4745997 + 2 n (-5451963 + 2937260 n))) g_*^3 r^(-4 k) \zeta^{12},\nonumber\\
a_4 &=& 2560 e (-6 + n) (-42 + 5 n) + 256 e^2 (-6 + n) (-42 + 5 n) - 
   5 (7770906 + 7 n (-2202012 + 964235 n))) g_*^4 \zeta^{10}, \nonumber\\
a_5&=&3 (-13589562 + 62592 e (10 + e) + 23072500 n - 14720 e (10 + e) n + 
   25 (-309417 + 32 e (10 + e)) n^2) g_*^5 \zeta^8\nonumber\\
a_6&=&5 (58176 e (10 + e) - 10560 e (10 + e) n + 
   25 (-70695 + 16 e (10 + e)) n^2 + 9 (-593417 + 854712 n)) g_*^6  \zeta^6,\nonumber\\
a_7&=& 25 (-398601 + 10116 e (10 + e) + 481842 n- 
   1260 e (10 + e) n + (-216 + 5 e) (266 + 5 e) n^2)g_*^7 \zeta^4\nonumber\\
a_8&=&-1500(-1080(n-1)+e(10+e)(5n-78))g_{*}^{8}\zeta^{2}, \nonumber\\
a_9&=& 22500g_{*}^{9}e(10+e). \nonumber
\end{eqnarray}

From the above coefficients, we can infer that $h(u)$ must have at least one zero. We note that for large $u$ we have that $h(u)$ is positive, but $h(0)$ is negative. Therefore, at some $u$, $h(u)$ must be null, and then, we must have a remnant. However, we can obtain the radius of the remnant only numerically.

Now, we turn to entropy. The first law of thermodynamics provides the following expression for the entropy
\begin{equation}
S(r_+)=\frac{1}{\sqrt{3}g_{*}}\int r_+^{1-2n}(g_* r_+^{2n}+\zeta^2)\sqrt{-g_{*}\lambda_{*}\left(1+\frac{\zeta^2}{g_{*}r_+^{2n}}\right)\left(-5+(5+e)\left(\frac{g_{*}}{g_{*}+\zeta^2 r_+^{-2n}}\right)^{3/2}+\frac{3}{8+10g_* r_+^{2n}\zeta^{-2}}\right)}d\,r_+.
\end{equation}

The integral above can not be solved analytically, and therefore we can not obtain an explicit form for the entropy in this case. Instead, let us consider an approximate form for the entropy for $\zeta<<1$, providing
\begin{equation}
S(r_+)=\frac{\sqrt{-eg_{*}\lambda_{*}}r_+^2}{2\sqrt{3}}-\frac{\sqrt{3}(-24+5e)\lambda_{*}\ln{(r_+)}\zeta^2}{20\sqrt{-eg_{*}\lambda_{*}}}+\frac{\sqrt{3}(1728+942e+25e^2)\lambda_{*}^{2}\zeta^4}{1600(-eg_{*}\lambda_{*})^{3/2}r_+^2}+O(\zeta^6),
\end{equation}
for the case $n=1$, and
\begin{equation}
S(r_+)=\frac{\sqrt{-eg_* \lambda_*} r_+^2}{2\sqrt{3}}+\frac{\sqrt{3}(5e-24)\lambda_* r^{2-2n}\zeta^2}{40\sqrt{-eg_* \lambda_*}(n-1)}+\frac{\sqrt{3}(1728+942e+25e^2)\lambda_* r^{2-4n}\zeta^4}{1600e(1-n)g_*\sqrt{-eg_*\lambda_*}}+O(\zeta^6).
\end{equation}

The  zero-order term in $\zeta$ is just the entropy obtained from the classical black string metric, while the $\zeta^2$ order term is just what we obtained from when we improved only the Newton's constant, that is, the logarithmic correction for $n=1$ and $r\propto 1/r^{2n}$ for $n>1$. Contrary to the case where only Newton's constant is running, we can see that a running cosmological constant introduces other correction terms beyond the corrections cited above. All these additional corrections depend on the inverse of the horizon radius and therefore become more relevant in the quantum regime, that is, in the final stage of the black string evaporation. 

\begin{figure}[!htpb]
    \centering
    \includegraphics{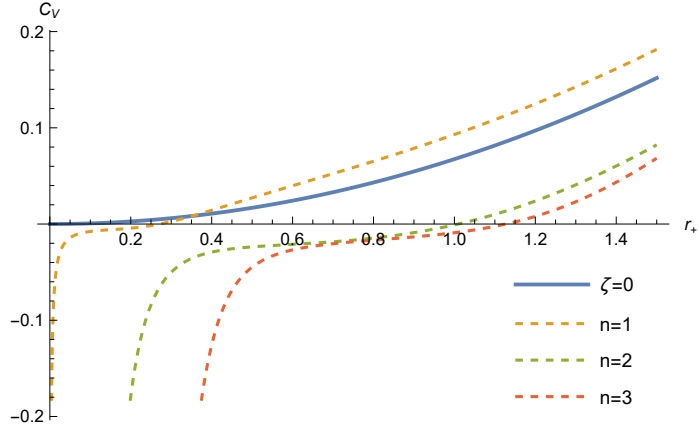}
    \caption{Heat capacity for the improved black string. We have set $\zeta=0.10$ and $e=-0.10$. The heat capacity obtained using the classical black string metric ($\zeta=0$) is plotted for comparison.}
    \label{l-heatcapacity}
\end{figure}

Finally, we check the thermodynamic stability of the improved black string solution in this case. To do this, we have to compute the heat capacity at constant volume. From the definition \eqref{def:heat-capacity} we can obtain the heat capacity, but the expression is too large to be reported here, and we show only the behavior of the heat capacity graphically. In fig (\ref{l-heatcapacity}) we plotted the heat capacity in this case for $\zeta=0.10$ and for the identifications with $n=1$, $n=2$ and $n=3$, in comparison with the usual expression of the heat capacity. We can, clearly see that the ASG effects produces a large deviation from the results with the classical black string metric. Again, the great difference is the presence of a remnant mass, which is not observed both in the usual case. The improved black string metric remains a stable system once is always positive, the region where becomes negative is forbidden, once in this region the temperature should be below the zero absolute. 

\section{Conclusion}

In this paper, for the first time, black strings were considered in the ASG scenario. To make the improvement, we have considered the quantum corrections from the asymptotic safety gravity. We implemented these corrections directly in the metric, via the improvement of Newton's constant and in the cosmological constant. We analyzed the thermodynamic properties of the improved black string solution in two cases: considering the cosmological constant is already fixed in a fixed point, and then only Newton's constant is improved, and the case where both Newton's constant and cosmological constants are running. In this work, we considered the ``naive'' identification $k \propto 1/r^n$ for the scale moment, whose corrections are guaranteed to disappear at the limit of great distances.

A notable result is that improving only the Newton's constant, the singularity is cured in the origin for the identifications with $n>1$, once the curvature scalars, such as the Ricci scalar and the Kretschmann scalar, are regular in the origin. However, by considering also a cosmological constant running, the problem of singularity remains in the black string solution.

All the thermodynamic quantities for the improved black string metric tends to their usual expression for large  values of the horizon radius and in the limit $\xi\rightarrow0$, where the ASG becomes negligible. Then, the corrections due to ASG becomes relevant only for small values of the horizon radius, as expected. 

The temperature for the improved solution in both cases has a small deviation from the linear behavior when the horizon radius approaches to zero. Besides that, we can observe that the temperature decreases when the value of the ASG parameter ($\xi$ in the first case and $\zeta$ in the second case) increases, that is, when the strength of the ASG corrections is increased. However, if only Newton's constant is improved, the temperature becomes null when the horizon radius becomes zero for the identification $n=1$, while for the cases $n>1$ the temperature is null for a non-null value of the horizon radius. This indicates that the identification with $n>1$ predicts the presence of a remnant mass in the final stage of evaporation. In contrast, if we improve both Newton's constant and the cosmological constant, the presence of a remnant mass is always observed, independently of the value of $n$, which is usually expected from quantum gravity corrections. 

For the corrected entropy, the ASG effects brings a logarithmic correction for the identification with $n=1$, which is a feature of quantum gravity corrections. For the cases $n>1$ we have only a correction of the type $1/r_+^{2n}$, emphasizing that the corrections becomes relevant only for small values of the horizon radius. However, differently in the case where the cosmological constant is fixed, the presence of a running cosmological constant introduces new corrections beyond the logarithmic term for $n=1$ and corrections with inverse of the distance for $n>1$. These corrections are inverse functions of the horizon radius, being more determinant for small distances. 

The heat capacity for the improved solution still remains positive in both cases and therefore we always have a stable system when the ASG corrections are considered. As expected, the deviation from the behavior for the usual case is observed only for small values of the horizon radius. Also, from the behavior of the heat capacity, we can see that the improved black string metric does not present phase transitions during its evaporation process, even when a running cosmological constant is considered. This is very different from what happens for the spherical case, such as the Schwarzschild anti-de Sitter black holes. 

Finally, we should point out that this manuscript is the first description of black strings in the ASG context, and therefore, many future studies can be done: consider other improvements and the different ways to implement them, study the particle trajectories in this improved spacetimes, theories with higher curvatures, and so on.  

\acknowledgments

The authors would like to thank Alexandra Elbakyan and sci-hub for removing all barriers in the way of science. We acknowledge the financial support provided by the Coordenação de Aperfeiçoamento de Pessoa de Nível Superior (CAPES), the Conselho Nacional de Desenvolvimento Científico e Tecnológico (CNPq) and Fundação Cearense de Apoio ao Desenvolvimento Científico e Tecnológico (FUNCAP) through PRONEM PNE0112-00085.01.00/16.

\end{document}